\documentclass[aps,prb,final,twocolumn,showpacs,superscriptaddress]{revtex4}
\usepackage{graphics}
\usepackage{epsfig}
\usepackage{units}
\usepackage{color}
\usepackage{amsmath}

\begin{document}

\title{Plasmons do not go that quantum}
\author{R. Carmina Monreal}
\email[Corresponding Author: ]{r.c.monreal@uam.es}
\affiliation{Departamento de F\'{\i}sica Te\'orica de la Materia Condensada and Condensed Matter Physics Centre, Universidad Aut\'onoma de Madrid, E-28049 Madrid, Spain.}
\author{Tomasz J. Antosiewicz}
\affiliation{Department of Applied Physics and Gothenburg Physics Centre, Chalmers University of Technology, SE-41296 G\"oteborg, Sweden}
\author{S. Peter Apell}
\affiliation{Department of Applied Physics and Gothenburg Physics Centre, Chalmers University of Technology, SE-41296 G\"oteborg, Sweden}


\begin{abstract}
We develop a theoretical model of the surface plasmon resonance of metallic nanospheres in the size range down to the single nanometer size. Within this model we explicitly show how different microscopic mechanisms, namely quantization due to size (QSE) and electron spill-out, affect the energy of the surface plasmon. We demonstrate, that electron spill-out effects, which can move the surface plasma energy both toward the red or the blue, can be comparable to or even stronger than QSE. Thus, depending on circumstances, QSE may only be observed for ultrasmall metal nanoparticles much closer to 1 nm in size than to 10 nm. Results presented herein are in quantitative agreement with recent published experimental results for Ag and Au.
\end{abstract}


\maketitle
During the last 20 years there has been considerable progress in both calculating and measuring the properties of small particles \cite{PRB_29_1558_ekardt, PRB_31_3486_puska, PRB_35_7325_beck, NL_12_429_townsend} as well as the corresponding efforts for clusters. Of particular interest is to monitor the region where cluster research meets and merges with the corresponding small particle research \cite{PRA_48_R1749_liebsch, PRL_74_1558_reiners, PRB_58_6748_yannouleas, TCA_116_514_cottancin}. Furthermore, in the last few years, the behavior of plasmon resonances in nanosystems has become a hot topic for its important applications in varied fields ranging from cancer therapy, nanophotonic devices, biosensing to catalysis. Interesting physical aspects of these resonances include, among others, investigation of the quantum regime of tunneling plasmonics \cite{Nature_491_547_savage} or what is an ultimate limit to electromagnetic enhancement in these systems \cite{Sci_337_1072_cira
 ci, OE_20_25201_apell}. In spite of all this effort, several aspects of the behavior of plasmon resonances in nanometer-sized metallic spheres are not fully understood. A particularly interesting example concerns the energy shift experienced by the surface plasmon as a function of the size and shape of the particle. There has been in the literature a variety of  experimental  reports of \emph{blueshifts} \cite{SSComm_13_153_smithard, ZPhysB_21_339_genzel, SSComm_16_113_smithard, PRB_33_2828_borensztein, TCA_116_514_cottancin, PRL_80_5105_lerme,  Nature_483_421_scholl, arXiv:1210.2535_raza} or \emph{redshifts} \cite{PRL_74_1558_reiners, PNAS_107_14530_peng} of these resonances as the particle dimensions are reduced,  which today may seem conflicting. There have, of course, been many attempts during the years to address this question theoretically \cite{SSComm_16_113_smithard, ZPhysA_270_17_lushnikov, PRB_11_2871_ruppin, SSComm_18_385_ascarelli, JOSA_66_449_ruppin, JPhysF_7_19
 35_boardman}. However, it is only recently that the experimental situation has become so well controlled that a realistic comparison can be made with theoretical model predictions.

The classical Mie electromagnetic theory \cite{AnnPhys_25_377_mie} is adequate for describing the optical properties of particles with a radius larger than 10 nm. Its main ingredient is the frequency dependent dielectric function $\epsilon(\omega)$ of the material and its most characteristic feature is the presence of a surface plasmon resonance. The resonance moves in energy as the particle size is changed, depending on the value of the radius $R$ with respect to the effective wavelength of the plasmon $c/\omega_{p}$, where $c$ is the speed of light and $\omega_{p}$ is the plasma frequency of the volume plasmon. The Mie theory gives rise to saturation of this energy at the value $\omega_{s}^{cl}=\omega_{p}/\sqrt{1+ 2\epsilon_m}$ for radii smaller than the wavelength, where $\epsilon_m$ is the dielectric constant of the medium surrounding the particle (assumed to be frequency independent).  When dealing with particles in the size-range 1-10 nm the surface starts to become mor
 e and more important compared with the bulk response and thus influences heavily collective modes localized in the surface region. Since electrons are able to spill out of the metal because of the finite potential barrier at the surface, the surface screening of an external perturbation is drastically changed from the Mie model assuming, as it does, a sharp interface between a particle and the surrounding medium. Whereas the classical theory is characterized by bulk dielectric properties and whence the length scale $R$ (the particle radius), the surface enters the electrodynamic response as a frequency dependent complex length scale $d_{r}(\omega)$ \cite{PS_26_113_apell}.  Its real part is a measure of the center of gravity of the screening charge induced in the particle by an external perturbation of frequency $\omega$ and  its imaginary part describes surface absorption. Reducing the radius below a few nanometers we move to a region where again the bulk becomes important i
 n a new fundamental way entering the quantum size effect region where the discrete energy level spectrum makes it possible to set up standing electron-hole pairs and quantum core plasmons \cite{PRB_29_1558_ekardt, NL_12_429_townsend}. The last are charge oscillations localized at the very center  of the particle because of the  quantum confinement of the electrons. These quantum size effects influence the particle response to a significant degree and compete with the effect coming from the surface.  For these ultra small particles sophisticated Time Dependent Density Functional Theory (TDDFT) and Time Dependent Local Density Approximation (TDLDA) calculations are available nowadays \cite{PRB_29_1558_ekardt, NL_12_429_townsend, TCA_116_514_cottancin, PRL_80_5105_lerme}. However, these become increasingly computationally demanding with increasing radius.

The purpose of this article is to clarify the role of the different microscopic mechanisms affecting the energy of the surface plasmon of metallic spheres as a function of their size, for diameters going down to a few nanometers. To do so we develop a theoretical model including surface spill-out and quantum size effects.  This model is based on  known concepts of surface screening and quantum size effects and successfully bridges cluster (sub 1 nanometer) research and Mie theory (greater than 10 nm) predicting results that agree with TDDFT calculations. Our work is triggered by recently available very detailed experiments \cite{Nature_483_421_scholl, arXiv:1210.2535_raza} which greatly improve the possibilities of checking basic electromagnetic response theory predictions with experimental findings. We note that the theoretical framework used in these papers does not appropriately include the major effect of the surface spill-out electrons thus giving the reader the impressi
 on that quantum effects start to enter the picture for diameters much larger than they actually do. We find that electron spill-out effects, which can move the surface plasma energy both toward the red or the blue, can be comparable to or even stronger than QSE.  We unambiguously show that taking surface screening and d-electrons properly into account the experimental surface plasmon energies of Ag and Au particles in a variety of host media can be reproduced down to a few nm sized particles and the discrepancy at smaller sizes can be accounted  for in a simple model of quantum size effects.

The electromagnetic response of a particle in the dipolar approximation can be described by its polarizability. The polarizability of a metal sphere of radius $R$ with a bulk dielectric constant $\epsilon(\omega)$ can be written as \cite{PS_26_113_apell, PRL_50_1316_apell, JPhysC_16_5729_apell}
\begin{equation}
\alpha(\omega)=R^3 \frac{\left(\epsilon(\omega)-\epsilon_m\right)\left(1-\frac{d_r(\omega)}{R}\right)}
{\epsilon(\omega)+2\epsilon_m+2\left(\epsilon(\omega)-\epsilon_m\right) \frac{d_r(\omega)}{R}},
\label{polarizability}
\end{equation}
where the length $d_r(\omega)$ is related to the charge density, $\delta \rho(r,\omega)$, induced by the external perturbation, by the formula
\begin{equation}
\frac{d_r(\omega)}{R}=\frac{ \int dr\; r(R-r)\delta\rho(r,\omega)}
{ \int dr\; r^2 \delta\rho(r,\omega)},
\label{dr-general}
\end{equation}
where $r$ is the radial coordinate and the integrals extend to the whole space. Notice, that for the Mie model, where the induced charge density is right at $r=R$, $d_{r}$ is identically zero and eq. (\ref{polarizability}) reduces to the classical Mie result when retardation is neglected. We will focus in this paper on the shift in energy of the surface plasma resonance, which can be obtained from the poles of the polarizability.  Then, from eq. (\ref{polarizability}), the frequency of the surface plasmon resonance, $\omega_s$, has to fulfill
\begin{equation}
\textrm{Re}\left[\epsilon(\omega_s)+2\epsilon_m+2(\epsilon(\omega_s)-\epsilon_m) \frac{d_r(\omega_s)}{R}\right]=0.
\label{eq-surface-plasmon}
\end{equation}
 
It has been known long since \cite{PRB_12_1319_fiebelman, PRB_32_6255_liebsch, PRB_36_7378_liebsch} that the calculated energies of the surface plasmons are very dependent on the shape of the barrier that confines the electronic system, even for semi-infinite metals. Infinite barrier models confining the electrons inside the metal produce frequencies higher than the classical value while realistic surface barriers can produce lower frequencies because electrons are allowed to spill-out of the metal, thus lowering  the plasma frequency. The same happens for nanospheres. If electrons are allowed to spill-out of the metal, the induced surface charge density has in general its main weight outside the surface with its center of gravity giving $\mathrm{Re}[d_r(\omega)]<0$, while a positive value of $\mathrm{Re}[d_r(\omega)]$  is obtained if electrons spill-in. From eq. (\ref {eq-surface-plasmon}) the \emph{color} of the shift of the surface plasmon frequency with respect to $\omega
 _{s}^{cl}$ is directly related to the sign of $\mathrm{Re}[d_r(\omega)]$: blue for positive and red for negative signs, respectively. Consequently, theoretical models in which the electrons in the nanoparticle are confined directly or indirectly by an infinite barrier \cite{PRB_32_7878_deAndres, Nature_483_421_scholl, arXiv:1210.2535_raza} can only produce blueshifts.

The above considerations refer to surface effects. Now we analyze quantum size effects (QSE) which are also present at the nanoscale size of interest here. These effects can be described approximately  by a dielectric constant similar to the one of a semiconductor: the discrete nature of the electronic states requires a minimum of energy to excite an electron, which is equivalent to having an energy gap. Inspired by the Penn dielectric function for semiconductors \cite{PhysRev_128_2093_penn}, we propose to use the following form of the real part of the dielectric function:
\begin{equation}
\epsilon(\omega)=1-\frac{\omega_{p}^{2}}{\omega^2-\Delta^2},
\label{eq-epsilon-QSE}
\end{equation}
where, as introduced by Gorkov and Eliashberg \cite{SovPhysJETP_21_940_gorkov}, the energy gap $\Delta$ scales with the particle radius as $\Delta=\omega_{p}R_{0}/R$. Then, using a simple box model, one obtains from their treatment $R_0\simeq 1.1a_0 \sqrt{r_s}$, where $r_s$ is the effective free-electron density parameter and $a_0$ is the Bohr radius. The existence of an energy gap influences the collective modes of the sphere. The classical bulk plasma frequency given by $\epsilon(\omega)=0$,  depends on the radius $R$ as
\begin{equation}
\omega_{p}(R)=\omega_p\sqrt{1+\left(\frac{R_0}{R}\right)^2},
\label{eq-wp}
\end{equation}
and, in the absence of any surface effect, the frequency of the surface plasmon changes to
\begin{equation}
\omega_{s}(R)=\omega_{s}^{cl}\sqrt{1+ (1+2\epsilon_m)\left(\frac{R_0}{R}\right)^2}.
\label{eq-wsp}
\end{equation}
That is, QSE will produce a blueshift of both surface and bulk plasmon energies. We notice that the relative shift is larger for the surface compared to the bulk plasmon.

Consequently, in a real situation with both quantum and surface barrier effects present, there is a competition between the blue- and the redshifts. Hence, in this work we apply the two just outlined models to calculate surface plasmon energies in small particles which will be compared to experimental results published in Refs. \cite{ZPhysD_12_471_charle, PRL_80_5105_lerme, Nature_483_421_scholl}. Since we will be dealing with the noble metals Ag and Au, we also need to include the screening due to the d-electrons by modifying eq. (\ref{eq-epsilon-QSE}) as
\begin{equation}
\epsilon(\omega)=\epsilon_d(\omega)-\frac{\omega_{p}^{2}}{\omega^2-\Delta^2},
\label{eq-epsilon-d}
\end{equation}
where $\epsilon_d(\omega)$ is obtained from experimental optical data \cite{PRB_6_4370_johnson}. This form of $\epsilon(\omega)$ is then introduced into eq. (\ref{eq-surface-plasmon}) to take into account quantum size effects and d-electrons on the surface plasmon frequency and we find that it has to fulfill
\begin{multline}
\omega_{s}^2=\omega_{p}^2 \left[\left(\frac{R_0}{R}\right)^2+ \right. \\
\left.\frac{1+2\textrm{Re}\left[\frac{d_r(\omega_s)}{R}\right]}{\epsilon_d(\omega_s)+2\epsilon_m+2(\epsilon_d(\omega_s)-\epsilon_m) \textrm{Re}\left[\frac{d_r(\omega_s)}{R}\right]}\right].
\label{eq-wsp2}
\end{multline}

\begin{figure}
\includegraphics[width=8.00cm]{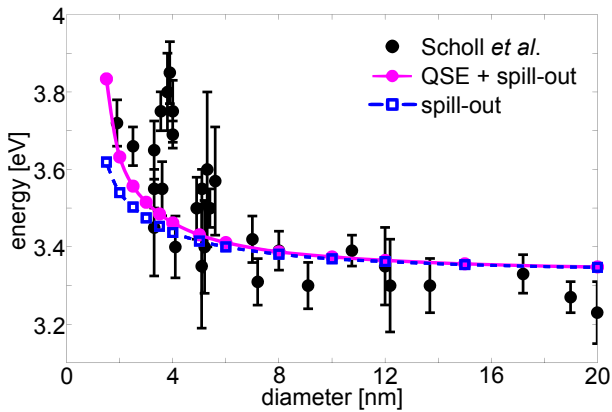}
\caption{Surface plasmon energy versus diameter for small Ag nanospheres. Points with error bars are experimentally extracted results from EELS measurements by Scholl et al. \cite{Nature_483_421_scholl}. The blue line is our calculation of the surface plasmon energy shift resulting from electron spill-out and the magenta line includes also the effects of size quantization. It is clear that the blue line gives a reasonable account for the data but the magenta line shows an even better agreement with the experiment. It has earlier been claimed that quantum effects start already at 10 nm however our calculations show that they do not dominate until the particles are about 10 times smaller. }
\label{fig::scholl}
\end{figure}

In the detailed treatment above we looked at the center of gravity of the induced density $d_r$. We now ask what is the influence of d-electrons and quantum size effects directly on $d_r$. We can do this to lowest order by looking to the corresponding quantity for a planar surface $d_{\perp}(\omega)$ because it has been shown \cite{PRB_29_1558_ekardt} that the induced charge density at the surface of a sphere is very similar to that of a planar surface down to a few nanometers in size. Also, we note that the d-electrons are very localized and therefore largely excluded from the surface region where the conduction electrons spill-out \cite{PRL_71_145_liebsch, PRB_48_11317_liebsch}. Hence, we will use the values of $d_{\perp}(\omega)$ calculated by Feibelman \cite{ProgSurfSci_12_287_feibelman} for a planar jellium surface of $r_s=3$ ($r_s=3.02$ and $r_s=3.01$ are the values giving the density of conduction electrons in Ag and Au, respectively). However, the screening of the d-e
 lectrons is very important in the bulk, in particular it affects the bulk plasma frequency by changing $\omega_p$ from its free-electron value of $9.1 $eV to $\omega_p^{*}= 3.81$ eV and $\omega_p^{*}= 5.99$ eV for Ag and Au, respectively.  Since $d_{\perp}$ only depends on the ratio $\omega/\omega_p$, we use for $d_r$ the values of $d_{\perp}$ with $\omega$ renormalized to $\omega_p^{*}$. Then we introduce the effects of d-electrons and of size quantization on $d_r$ trough $\omega_p^{*}(R)$.

We apply eq. (\ref{eq-wsp2}) to calculate the energies of the surface plasmon resonances of three different metals: Ag, Au, and a pure free-electron-metal of $r_s=3$ . These metals have in common their close values of the effective free-electron density parameter $r_s$. However, the screening caused by the d-electrons is very different in Ag, Au, and a free-electron metal at the typical frequencies of the surface plasmons. This is a crucial physical property for making the resonances of metal nanoparticles to shift to different colors.

\begin{figure}
\includegraphics[width=8.00cm]{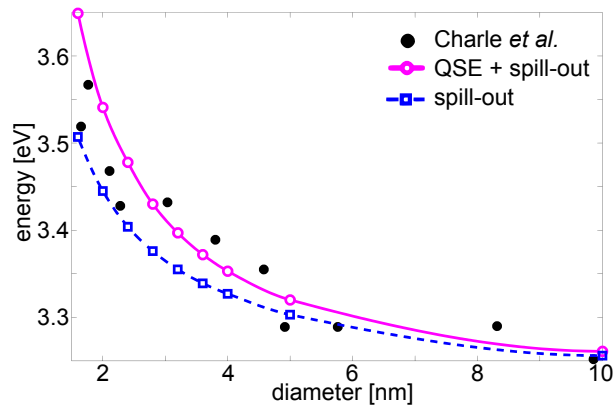}
\caption{Surface plasmon energy shift for Ag particles embedded in a solid argon matrix. Full dots are the experimental data of Charl\'{e} et al. \cite{ZPhysD_12_471_charle}. The blue line shows our results from electron spill-out and the magenta line includes also the effects of size quantization.}
\label{fig::Ag_on_Ar}
\end{figure}

Figure \ref{fig::scholl} shows the calculated frequencies of the surface plasmons of Ag nanospheres as a function of the diameter, together with the experimental results by Scholl et al. for Ag spheres deposited on carbon foils obtained using electron energy loss spectroscopy (EELS) \cite{Nature_483_421_scholl} (points with error bars). In our calculation we use  $\epsilon_m=1.5$, as an average of the dielectric constant of the medium surrounding the nanospheres, which reproduces the measured surface plasmon energies at the largest sizes. Even though there is large scatter in the experimental data there is a clear trend of increasing plasma frequency as the particles get smaller. The blue line is our calculation of the surface plasmon energy shift resulting from electron spill-out. Usually this spill-out gives a decrease in surface plasmon energy but for Ag the screening of the d-electrons can in some circumstances push the surface plasmon energy into a region where the dynam
 ical $\mathrm{Re}[d_r(\omega_s)]$ is positive and gives a blueshift \cite{PRB_33_2828_borensztein}. It is clear that the blue line gives a reasonable account for the data except at the very smallest particles of a size 1-2 nm. In this region we do expect effects from the finite size of the particles, introducing discrete energy levels and a gap at the Fermi level. Accounting for this as indicated by eq. (\ref{eq-epsilon-d}) and using the simple box estimation of $R_0$, $R_0= 1.1a_0 \sqrt{r_s}$, eq. (\ref{eq-wsp2}) results in the magenta line yielding an even better agreement with the measured data. We would like to stress that the only free parameter of this calculation is $\epsilon_m$, which, however, is constrained in its value, since it has to yield the asymptotically measured plasmon energies. Furthermore, the resulting value is very close to the average of air and carbon used in the experiments as support.

The good agreement between theory and experiment found for Ag particles on carbon foils is not fortuitous.  We show this by analyzing other experiments in which Ag and also Au nanospheres are embedded in solid matrixes. Figure \ref{fig::Ag_on_Ar} shows the calculated frequencies of the surface plasmons of Ag  together with the experimental results by Charl\'e et al. \cite {ZPhysD_12_471_charle} for Ag spheres embedded in a solid argon matrix. We have used  $\epsilon_m=1.85$ which is very close to the value $\epsilon_m=1.75$ quoted in  \cite{ZPhysD_12_471_charle} for solid argon. The blueshift of the resonance is explained exactly on the same basis as the results of Fig. \ref{fig::scholl} and again this simple model is able to reproduce the experimental data on a quantitative level. Notice that the maximum value of the shift measured (and calculated) is approximately 0.3~eV  while it was 0.5~eV in  Fig. \ref{fig::scholl}. This is due to the fact that as $\epsilon_m$ increases,
  $\omega_{s}^{cl}$ decreases thus moving to the a region of frequencies where the dynamical $\mathrm{Re}[d_r(\omega)]$ is still positive but smaller than in the previous case.

\begin{figure}
\includegraphics[width=8.00cm]{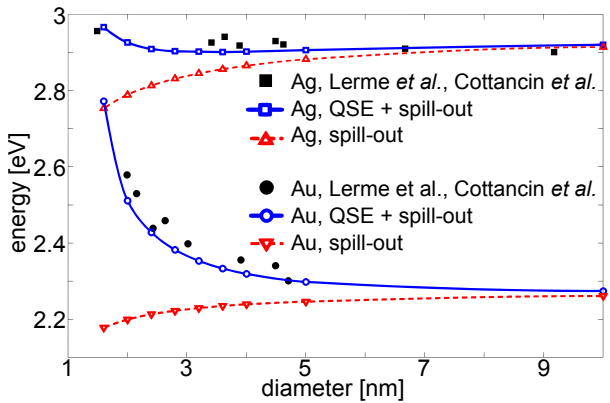}
\caption{Surface plasmon energy positions of Ag and Au particles embedded in porous alumina. Full black symbols without a line represent the experimental data for Ag (squares) and Au (dots) from works of Lerm\'{e} et al. \cite{PRL_80_5105_lerme} and Cottancin et al. \cite{TCA_116_514_cottancin}. The red lines represent our results from electron spill-out and the blue lines also include QSE. We see here that for Ag (top) we are in a redshifted spill-out region. However for very small particles the quantum size effects contributes a shift in the opposite way which even upsets the redshift. In the bottom part of the figure we illustrate that our treatment also works for other metal particles, in this case gold.}
\label{fig::lerme}
\end{figure}

Figure \ref{fig::lerme} shows the calculated frequencies of the surface plasmons of Ag and Au nanospheres together with the experimental results by Cottancin et al. \cite{TCA_116_514_cottancin} and Lerm\'e et al. \cite {PRL_80_5105_lerme} for Ag and Au spheres embedded in alumina (black symbols). In all the calculations we have used the value $\epsilon_m=2.8$ appropriate for porous alumina  \cite {TCA_116_514_cottancin}. As in Figs. \ref{fig::scholl} and \ref{fig::Ag_on_Ar}, we show results considering only surface effects and including effects of size quantization as well. 

First, we discuss the case of Ag. Notice that the relevant surface plasmon energies now are pushed down below 3 eV as compared to Figs. \ref{fig::scholl} and \ref{fig::Ag_on_Ar} due to the different dielectric properties of alumina as compared to carbon/vacuum and solid argon, respectively. Here we find that surface effects cause a \emph{redshift} of the resonance. This is because, as we have anticipated, an increased value of $\epsilon_m$ has moved the classical Mie frequency to the region of negative values of $\textrm{Re}[d_r]$ thus changing the sign of the shift. Then for Ag we are usually in a region where the spill-out effect can go both red and blue for a small change of frequency. In the present case the effects of size quantization compensate the redshift in the way of keeping the energy of the surface plasmon almost constant as the size decreases, as seen in the experiment.

In the case of Au, we also obtain that surface effects would shift the frequency toward the red but now the blueshift resulting from size quantization is stronger, producing the net blueshift shown in the figure. For Au we use an $R_0$ which is 40\% larger than for Ag. A possible reason for this is that, even though electrons in the d-bands will in principle also experience size effects, these should be smaller for Ag with deeper d-bands than for Au. Notice in eq. (\ref{eq-wsp2}) how the relative surface versus quantum size effects depend on $\epsilon_d(\omega)$. This effect is illustrated in the calculations we present next, for which $\epsilon_d(\omega)=1$.

\begin{figure}
\includegraphics[width=8.00cm]{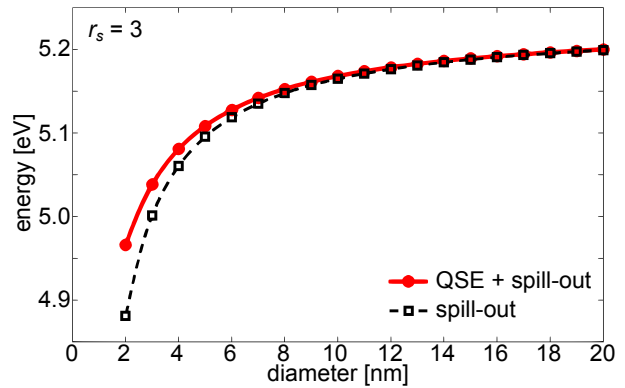}
\caption{Surface plasmon energy positions of free-electron-metal particles of $r_s=3$ in vacuum. The redshift produced by surface spill-out cannot be overcome by effects of size quantization as in the case of Au and Ag and the surface plasmon moves to lower energies with decreasing size.}
\label{fig::rs}
\end{figure}

Figure \ref{fig::rs} shows our results for free-electron-metal nanospheres of $r_s=3$ in vacuum, using as before the simple box estimate for $R_0$. This is an example where the surface effects are practically responsible for the strong redshift of the surface resonance, size quantization effects being unable to compensate for it. We would like to mention that sophisticated  calculations for this system have produced redshifts of the surface resonances \cite{PRB_29_1558_ekardt, NL_12_429_townsend}.  The value we find for a radius of $R=0.74$~nm is $\omega_s=0.94\omega_s^{cl}$ which is close to the value $\omega_s=0.91\omega_s^{cl}$ obtained by Townsend and Bryant \cite{NL_12_429_townsend} using TDDFT. Moreover, a net redshift of these resonances has been measured for nanospheres of the typical free-electron metals Na and K in Ref. \cite{PRL_74_1558_reiners} and the same behavior is expected for other such metals as Al and Mg.

We have developed a model theory that is in good agreement with existing sophisticated calculations only feasible for very small systems and yields excellent quantitative agreement with a variety of available experimental results for metallic spheres of nanometric size. We find that the effects of the spill-out of the electronic charge at the surface of the nanoparticle, which can move the surface plasma energy toward the red or the blue depending on the circumstances, are never negligible in comparison to effects of size quantization which are also present in these small systems. We thus conclude, that the surface plasmon resonance in small particles does not go that quantum as recent experiments seem to suggest \cite{Nature_483_421_scholl, arXiv:1210.2535_raza}: instead of 10 nm we would expect to see effects unambiguously rather at 1 nm and below.

\acknowledgments
We thank T. L\'opez-R\'{\i}os for useful discussions. R. C. Monreal acknowledges financial support from the Spanish Mineco, project FIS2011-26516. T.J.A. and S.P.A acknowledge financial support from the Swedish Foundation for Strategic Research via the Functional Electromagnetic Metamaterials project SSF RMA 08. 


\end{document}